\documentclass[10pt,pra,aps,superscriptaddress,twocolumn,longbibliography,showpacs]{revtex4-1}
\usepackage{amssymb,amsmath}
\usepackage{graphicx}
\usepackage{color}
\usepackage{natbib}
\usepackage{hyperref}
\usepackage[utf8]{inputenc}

\begin{document}

\newcommand{\e}{{\rm e}}
\newcommand{\rmi}{{\rm i}}
\renewcommand{\Im}{\mathop\mathrm{Im}\nolimits}
\renewcommand{\Re}{\mathop\mathrm{Re}\nolimits}
\newcommand{\red}[1]{{\color{red}#1}}
\newcommand{\blue}[1]{{\color{blue}#1}}
\newcommand{\commentSasha}[1]{{\color{blue}{\bf Sasha:}\it #1}}
\newcommand{\commentMaxim}[1]{{\color{red}{\it ~Maxim:~}\tt #1}} 

\newcommand{\refPR}[1]{[\onlinecite{#1}]}
\newcommand{\cra}[1]{\hat{a}^{\dag}_{#1}}  
\newcommand{\ana}[1]{\hat{a}_{#1}^{\vphantom{\dag}}}         
\newcommand{\num}[1]{\hat{n}_{#1}}         
\newcommand{\bra}[1]{\left|#1\right>}      
\newcommand{\eps}{\varepsilon}      
\newcommand{\om}{\omega}      
\newcommand{\kap}{\varkappa}      

\newcommand{\spm}[1]{#1^{(\pm)}}
\newcommand{\skvk}[2]{\left<#1\left|\frac{\partial #2}{\partial k}\right.\right>} 
\newcommand{\skvv}[2]{\left<#1\left|#2\right.\right>}
\newcommand{\df}[1]{\frac{\partial #1}{\partial k}}
\newcommand{\ds}[1]{\partial #1/\partial k}

\title{Interaction-induced two-photon edge states in extended Hubbard model \\ realized in a cavity array}
\author{Maxim~A.~Gorlach}
\affiliation{ITMO University, Saint Petersburg 197101, Russia}
\email{Maxim.Gorlach.blr@gmail.com}
\author{Alexander N. Poddubny}
\affiliation{ITMO University, Saint Petersburg 197101, Russia}
\affiliation{Ioffe Institute, Saint Petersburg 194021, Russia}
\email{poddubny@coherent.ioffe.ru}

\begin{abstract}
We study  theoretically two-photon states in a periodic array of coupled cavities  with both on-site and nonlocal Kerr-type nonlinearities. In the absence of nonlinearity the structure is  topologically trivial and possesses no edge states. The interplay of two nonlinear interaction mechanisms described by the extended Hubbard model facilitates formation of edge states of bound photon pairs. Numerical and exact analytical results for the two-photon wave functions are presented. 
Our findings thus shed light onto the edge states of composite particles and their localization properties.
\end{abstract}

\maketitle

\section{Introduction}\label{sec:Intro}

Photonic edge states are currently under active research in the context of topological photonics~\cite{Lu2014} due to their prospects for disorder-robust routing of light on a chip. So far most of the studies have been  focused  on the regime of linear classical optics~\cite{Lu2014,Lu2016}. The emerging nonlinear topological photonics  promises dynamical tuning of light propagation pathways, with  existing studies being only theoretical so far~\cite{Solnyshkov2016,Khanikaev2016,Leykam2016,Nalitov2016,Gulevich2016}. The ultimate limit of tunability is reached in the quantum regime  with several interacting photons~\cite{Muller2015}. Entangled multiphoton states  are useful for quantum  information and metrology applications~\cite{Afek,Solntsev2014}. Up to now most of the studies have been focused on bulk states of bound particle pairs and triplets  \cite{Winkler,Valiente,Valiente-JOP,Wang2010,Tangpanitanon2016}.
Here we investigate the effect of photon-photon interactions on the photon pair edge states.
 
   One of the simplest models incorporating interactions is the Bose-Hubbard model with the local interaction between the particles~\cite{Hartmann-LPR}. Despite its conceptual simplicity this model is capable to capture rich physics including the formation of repulsively bound  pairs of particles (doublons)~\cite{Winkler,Valiente,Pinto} that have been observed in optical lattices of cold atoms~\cite{Winkler}.
However, the edge states of bound photon pairs are not present in a periodic array of identical resonators with on-site Bose-Hubbard interaction~\cite{Flach} even though they can be enabled by an edge defect~\cite{Longhi,Zhang2012,Zhang2013}. Another way to implement the edge states of bound two-photon quasiparticles is to use a periodic array with two coupled cavities per unit cell with the different inter- and intracell tunneling amplitudes~\cite{Gorlach,CarusottoPRA}. Such structure is known to be topologically nontrivial even in the single particle scenario~\cite{Shen}.

In the present work we explore a different option to induce doublon edge states. Specifically, we employ an  extended Hubbard model~\cite{Dutta} that takes into account the nonlocal  interaction of the photons in the neighboring cavities. As it was shown previously~\cite{Valiente-JOP,Wang2010,Carusotto-2017}, such type of nonlinearity can lead to the two types of bulk doublon modes: (i) photons co-localized in the same cavity due to the usual on-site nonlinearity and (ii) photons co-localized in the neighboring cavities due to the nonlocal nonlinearity specific to the extended Hubbard model. 

\begin{figure}[b]
   \begin{center}
   \includegraphics[width=0.6\linewidth]{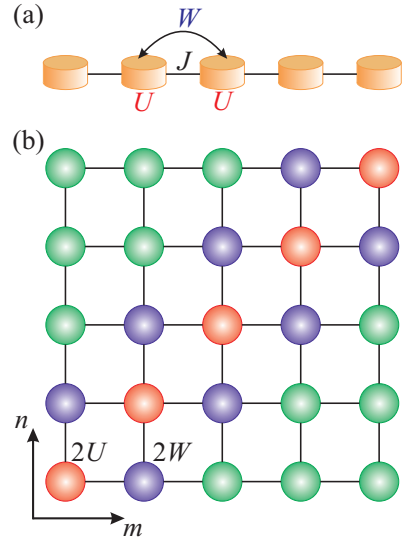}
   \caption{(a) Array of identical coupled resonators described by the extended Hubbard model. (b) Illustration of the equivalent two-dimensional problem.} 
   \label{fig:TwoDimensional}
   \end{center}
   \end{figure}

The considered one-dimensional array of coupled  cavities arranged in a simple periodic lattice [Fig.~\ref{fig:TwoDimensional}(a)] is described by the Hamiltonian
%

\begin{align}\label{ExtHubbard}
\hat{H}=&\omega_0\,\sum\limits_m\,\hat{n}_m+U\,\sum\limits_m\,\hat{n}_m\,(\hat{n}_m-1)\\\nonumber
&-J\,\sum\limits_m\,\left(\cra{m}\ana{m+1}+\cra{m+1}\ana{m}\right)+2W\,\sum\limits_m\,\hat{n}_m\,\hat{n}_{m+1}\:,
\end{align}
where the term $\propto U$ describes the on-site nonlinearity; the term  $\propto J$ describes the coupling between the neighboring cavities, and the $W$-term is responsible for the nonlocal interaction between the photons in  the neighboring cavities.
We demonstrate that even though the corresponding single-photon problem is topologically trivial and there are no defects in the system, it can still support the edge states of bound photon pairs. We also reveal  the biexponential character of the  doublon edge state localization reflecting the composite nature of the quasi-particles.

The rest of the paper is organized as follows. In Sec.~\ref{sec:Doublons} we revisit the bound photon pairs dispersion and explore the phenomenon of doublon  collapse into noninteracting photons in more detail. Section~\ref{sec:EffHam} gives a qualitative reasoning of the doublon edge state emergence. Section~\ref{sec:Edge} contains the exact analytical derivation of doublon edge states. Section~\ref{sec:Concl} is reserved for conclusions. Appendix describes the derivation of the doublon effective Hamiltonian.

\section{Formation and collapse of bound photon pairs}\label{sec:Doublons}

We start with the description of doublon states in an infinite chain. Since the particle number operator $\hat{N}=\sum_{m}\,\hat{n}_m$ commutes with the Hamiltonian Eq.~\eqref{ExtHubbard}, the number of photons is conserved and we can search for the two-photon wave function in the form
\begin{equation}\label{TwoPhotWave}
\bra{\psi}=\sum\limits_m\,\beta_{mm}\,\sqrt{2}\,\bra{2_m}+\sum\limits_{m\not= n}\,\beta_{mn}\,\bra{1_m\,1_n}
\end{equation}
with $\beta_{mn}=\beta_{nm}$. Combining Eq.~\eqref{TwoPhotWave} and the eigenvalue equation $\hat{H}\bra{\psi}=(\eps+2\,\omega_0)\,\bra{\psi}$ with the Hamiltonian  Eq.~\eqref{ExtHubbard} we obtain the following system of equations:
\begin{align}
(\eps-2U)\,&\beta_{mm}=-2J\,\beta_{m,m+1}-2J\,\beta_{m-1,m}\:,\label{Sys1}\\
(\eps-2W)\,&\beta_{m,m+1}=-J\,\bigl[\beta_{m+1,m+1}\label{Sys2}\\+
&\beta_{m,m+2}+\beta_{mm}+\beta_{m-1,m+1}\bigr]\:,\notag\\
\eps\,&\beta_{mn}=-J\,\bigl[\beta_{m+1,n}+\beta_{m,n+1)}\label{Sys3}\\\notag&+
\beta_{m-1,n}+\beta_{m,n-1}\bigr]\:,\mspace{15mu} (|m-n|\geq 2)\:.
\end{align}
The system of equations Eqs.~\eqref{Sys1}--\eqref{Sys3} can be interpreted as an effective two-dimensional one-particle problem~\cite{Longhi} illustrated in Fig.~\ref{fig:TwoDimensional}(b). The  two-dimensional structure consists of a square lattice of coupled cavities with the  main diagonal   ($m=n$) detuned by the energy $2U$ and the second  diagonals ($m=n\pm 1$)  detuned by $2W$. Quite importantly, the structure in Fig.~\ref{fig:TwoDimensional}(b) can be fully emulated classically  by the evanescently coupled array of waveguides, see e.g. the recent experiment Ref.~\cite{Mukherjee}.

Employing the translational invariance of the lattice we introduce the Bloch wave number $k$ characterizing the  motion of a photon pair as a whole:
\begin{equation}\label{TranslInv}
\beta_{mn}=\gamma_{n-m}\,e^{ik(m+n)/2}\:.
\end{equation}
The coefficients $\gamma_n\equiv\gamma_{-n}$ describe the contribution from the photon pairs separated by $n$ resonators. These coefficients satisfy the following system of equations:
\begin{align}
(&\eps-2U)\,\gamma_0=-2\,t\,\gamma_1\:,\label{Gam0}\\
(&\eps-2W)\,\gamma_1=-t\,\left(\gamma_0+\gamma_2\right)\:,\label{Gam1}\\
&\eps\,\gamma_n=-t\,\left(\gamma_{n-1}+\gamma_{n+1}\right),\mspace{15mu} (n\geq 2)\:,\label{Gam2}
\end{align}
where $t=2J\,\cos(k/2)$. We seek the solution for Eqs.~\eqref{Gam0}--\eqref{Gam2} in the form
\begin{equation}\label{GamAnsatz}
\gamma_n=C\,e^{i\kap(n-1)/2}\mspace{15mu} (n\geq 1)\:,
\end{equation}
where the wave number $\kap$ describes the relative motion of the photons.
The dispersion law for two-photon excitations reads
\begin{equation}\label{DispLaw1}
\begin{split}
\eps=-2t\,\cos(\kap/2)=-4J\,\cos(k/2)\,\cos(\kap/2)\\=
-2J(\cos k_1+\cos k_2)\:,
\end{split}
\end{equation}
where $k_{1,2}=(k\mp \kap)/2$. For real values of  $k$ and $\kap$ in Eq.~\eqref{DispLaw1}, the dispersion law describes two quasi-independent photons with the energy equal to a sum of single-photon energies. Both wave numbers $k_{1,2}$ for quasi-independent photons span the range $[-\pi,\pi]$ and thus $k=k_1+k_2$ varies in the range $[-2\pi,2\pi]$. Doublon solutions are described by the coefficients $\gamma_n$  that decay with the increase of $n$, i.e. $\Im\kap>0$  in Eq.~\eqref{GamAnsatz}. Contrary to the quasi-independent photon pairs, for doublons the center-of-mass wave number $k$ varies in the range from $-\pi$ to $\pi$. Equations~\eqref{Gam0}--\eqref{Gam2} yield
the algebraic equation for $z=e^{i\kap/2}$
\begin{equation}\label{KappaEquation}
2Wt\,z^3+(4UW-t^2)\,z^2+2(U+W)\,t\,z+t^2=0\:
\end{equation}
determining  the energy of the doublons $\eps=-t(z+z^{-1})$ for the solutions with $|z|<1$.
\begin{figure}[t]
\begin{center}
\includegraphics[width=1.0\linewidth]{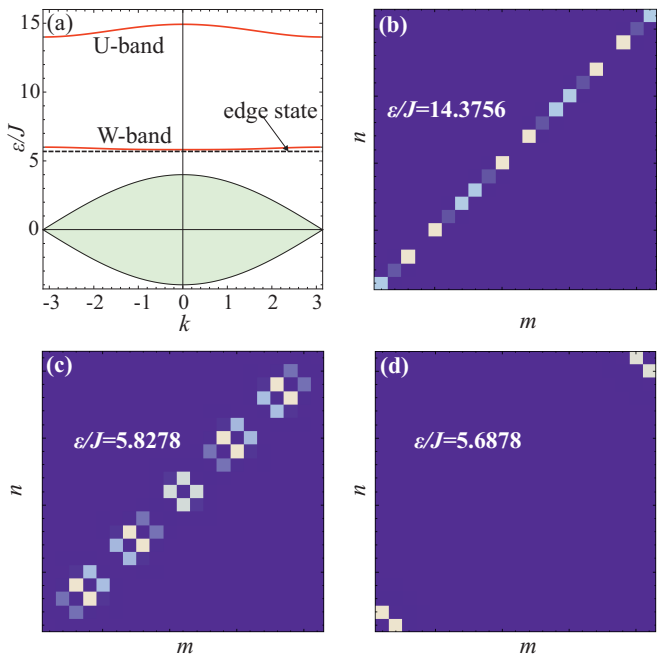}
\caption{(a) Calculated doublon dispersion and the continuum of quasi-independent photon states for $U/J=7$, $W/J=3$. Dashed line indicates the energy of the doublon edge state. (b-d) Probability distributions calculated for for a finite chain composed of $N=21$ resonator: (b) doublon U-band; (c) doublon W-band; (d) doubly degenerate doublon edge state.} 
\label{fig:DoublonDispersion}
\end{center}
\end{figure}
The doublon dispersion law $\eps(k)$ can be readily found from Eq.~\eqref{KappaEquation} and it is illustrated in Fig.~\ref{fig:DoublonDispersion}(a). Calculation shows that there exist two doublon bands in the system. The first band corresponds to the two photons  co-localized in the same cavity [Fig.~\ref{fig:DoublonDispersion}(b)] and the second one corresponds to the two photons co-localized in the neighboring cavities [Fig.~\ref{fig:DoublonDispersion}(c)]. As will be shown later, in the finite structure the  latter band can give rise to the  doublon edge states localized at  the array  edges [Fig.~\ref{fig:DoublonDispersion}(d)].

In order to better understand the system behavior, we study certain limiting cases of Eq.~\eqref{KappaEquation} with the analytical solution available. Namely, if the nonlocal interaction is absent, $W=0$, we return back to the simple chain of nonlinear  cavities possessing only one doublon band with the energy~\cite{Winkler}
\begin{equation}\label{EnergySim}
\eps=2\,\text{sign}\,U\,\sqrt{U^2+4J^2\,\cos^2(k/2)}\:.
\end{equation}
Analytical solutions can be also found for the doublon wave numbers $k$ close to the  boundaries of the first Brillouin zone $k=\pm\pi$, where $t(k)=0$.  The solutions of Eq.~\eqref{KappaEquation} up  to the leading order in $t$ are given by $z_1=-t/(2U)$ and $z_2=-t/(2W)$ and correspond to the energies $\eps_1=2U$ and $\eps_2=2W$. These expressions provide exact energies of doublon states in the limit $k=\pi$. On that basis, from now on we term the two doublon bands as the $U$-band and the $W$-band, respectively.

\begin{figure}[t]
\begin{center}
\includegraphics[width=0.95\linewidth]{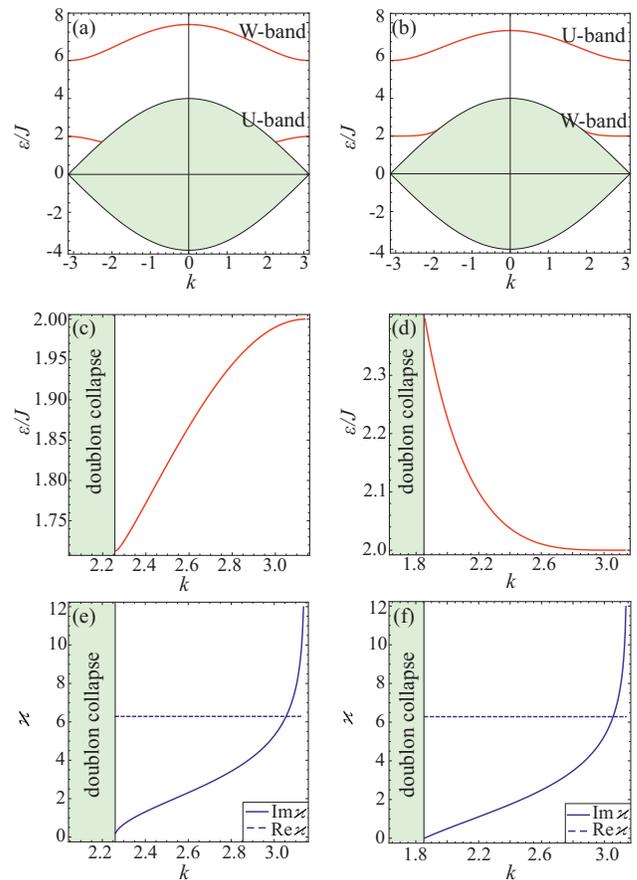}
\caption{(a,b) Doublon dispersion and the continuum of quasi-independent photon states. (c,d) Enlarged fragment of the doublon dispersion curve in the region of collapse. (e,f) Real and imaginary parts of the wave number $\kap$ describing  photons co-localization in the region of collapse. Calculations are performed for (a,c,e) $U/J=1$, $W/J=3$; (b,d,f) $U/J=3$, $W/J=1$.} 
\label{fig:Collapse}
\end{center}
\end{figure}

One more remarkable feature of the system is the phenomenon of doublon collapse into the pairs of quasi-independent photons. Namely, for some parameter values the dispersion branch of the doublon {can intersect with the continuum} of quasi-independent photon states as illustrated in Fig.~\ref{fig:Collapse}(a-d). Both $U$- and $W$-bands are subject to collapse [Fig.~\ref{fig:Collapse}(a,c) and Fig.~\ref{fig:Collapse}(b,d), respectively]. The doublon is always stable for wave vectors at the boundary of the first Brillouin zone, whereas sufficiently long-wavelength  doublons can decay. In the case of collapse  the imaginary part of the doublon localization parameter $\kap$ turns to zero [Fig.~\ref{fig:Collapse}(e,f)] which means that the photons are no longer confined to each other. Such bifurcation in the doublon dispersion can have an interesting impact on the behavior of the system driven by the external field resulting in the sudden death of doublon Bloch oscillations~\cite{Lin}.

We now examine the number of stable doublon bands as a function of the local and non-local photon interaction parameters $U/J$ and $W/J$. To this end we notice that the condition for the doublon collapse reads $z=e^{i\kap/2}=\pm 1$ which means $\Im\kap=0$ and $\Re\kap=0$ or $\Re\kap=\pi$  [see e.g. Fig.~\ref{fig:Collapse}(e,f)]. These conditions immediately yield the values of the  the parameter $W$ corresponding to the collapse:
\begin{gather}
W_1(U)=\frac{U\,J}{U-2J}\:,\label{CollapseCond}\\
\text{or}  \mspace{20mu}W_2(U)=-\frac{U\,J}{U+2J}\:.\label{CollapseCond2}
\end{gather}
The stability regions of the doublon bands calculated from Eqs.~\eqref{CollapseCond}, \eqref{CollapseCond2} are illustrated in Fig.~\ref{fig:FullDiagram}(a).

\section{Effective one-dimensional model for the  doublon edge states}\label{sec:EffHam}
\begin{figure}[b]
\begin{center}
\includegraphics[width=0.8\linewidth]{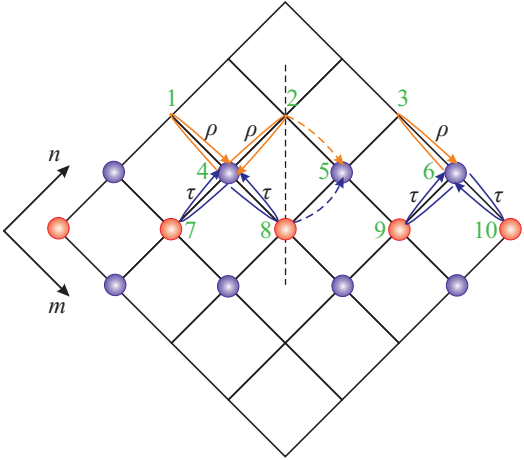}
\caption{Doublon tunneling pathways and qualitative explanation of the doublon edge states emergence.} 
\label{fig:Pathways}
\end{center}
\end{figure}

In this section we provide a qualitative explanation of the doublon edge state emergence for the $W$-band in the regime of strong interaction $W,U\gg J$.  If $J=0$, the doublons are confined to the neighboring cavities, $|m-n|= 1$ and have the energy $2W$. The finite values of $J$ enable tunneling of the doublons as a whole as well as shift their energies. The general form of the effective Hamiltonian for the doublon $W$-band is 
\begin{equation}\label{EffHamGeneral}
\hat{H}_{\rm{eff}}=\sum\limits_{l=1}^{N-1}\,\left(2\,W+\delta_l\right)\,\hat{d}_l^{\dag}\,\hat{d}^{\vphantom{\dag}}_l+f\,\sum\limits_{l=1}^{N-2}\,\left[\hat{d}_l^{\dag}\,\hat{d}^{\vphantom{\dag}}_{l+1}+\text{H.c.}\right]\:,
\end{equation}
where $\hat{d}_l=\ana{l}\,\ana{l+1}$, $f$ is the effective tunneling constant for doublons, and $\delta_l$ is the self-induced nonlinear blueshift of the eigenfrequency.   We find the values of $f$ and $\delta_l$ by means of the second-order perturbation theory in $J/U$, $J/W$ which corresponds to including the two-step photon tunneling pathways shown in Fig.~\ref{fig:Pathways}~\cite{Flach,Gorlach}.

The blueshift for all the bulk sites $\delta_2=\delta_3=\dots=\delta_{N-2}=\delta_{\rm{B}}$ is determined by the contributions from the four closed two-step paths $4-1-4$, $4-2-4$, $4-7-4$, $4-8-4$ (Fig.~\ref{fig:Pathways}). If the amplitude corresponding to the first two paths is equal to $\rho$ and the amplitude corresponding to the latter two paths is equal to $\tau$ then $\delta_{\rm{B}}=2\tau+2\rho$. The blueshift of the edge site $6$ in Fig.~\ref{fig:Pathways} is determined only by the three contributions and thus $\delta_{\rm{E}}=2\tau+\rho$. The tunneling amplitude $f$ is also determined by the two-step paths, e.g. $4-2-5$ and $4-8-5$. Using the mirror symmetry of the system with respect to the line $2-8$ these paths can be mapped onto two closed paths $4-2-4$ and $4-8-4$. Therefore, the tunneling amplitude has two contributions, $f=\tau+\rho$.

Thus, the problem of doublon edge states can be effectively mapped onto the problem of edge states in a one-dimensional array of identical cavities  with the edge detuned by the energy $\delta_{\rm E}-\delta_{\rm B}$. 
The edge states are present  provided that the detuning exceeds the tunneling constant~\cite{Gorlach}, i.e. $|\delta_{\rm{E}}-\delta_{\rm{B}}|>|f|$ or $|\rho|>|\tau+\rho|$. As it is shown in Appendix, $\tau=2J^2/(\eps-2U)\approx J^2/(W-U)$ and $\rho=J^2/\eps\approx J^2/2W$. The resulting condition for the doublon edge state emergence has the form
\begin{gather}
|W|<|U|/2\:,\mspace{15mu} \text{and} \label{Cond}\\
\text{sign}\,W=\text{sign}\,U\:.\label{Cond2}
\end{gather}

A similar analysis shows that no doublon edge states are associated with the $U$-band, the details are provided in Appendix.

\section{Doublon edge states}\label{sec:Edge}

Next we provide an exact analytical derivation of the doublon edge states in a semi-infinite array valid for arbitrary photon-photon interaction strength. If the array starts from the resonator with the number $1$, the system of equations for $\beta_{mn}$ coefficients reads
\begin{align}
& (\eps-2U)\,\beta_{mm}=-2J\,\beta_{m,m+1}-2J\,\beta_{m-1,m}\:,\label{NSys1}\\
& (\eps-2W)\,\beta_{m,m+1}=-J\,[\beta_{m+1,m+1}\label{NSys2}\\
&\quad + \beta_{m(m+2)}+\beta_{mm}+\beta_{m-1,m+1}]\:,\nonumber\\
& \eps\,\beta_{mn}=-J\,[\beta_{m+1,n}+\beta_{m,n+1}\label{NSys3}\\&\quad +\notag
\beta_{m-1,n}+\beta_{m,n-1}],\quad (|m-n|\geq 2)\:.\\\
& (\eps-2U)\,\beta_{11}=-2J\,\beta_{12}\:,\label{NSys4}\\
& (\eps-2W)\,\beta_{12}=-J\,[\beta_{11}+\beta_{22}+\beta_{13}]\:,\label{NSys5}\\
& \eps\,\beta_{1n}=-J\,[\beta_{1,n-1}+\beta_{1,n+1}+\beta_{2n}]\:.\label{NSys6}
\end{align}
Boundary conditions Eqs.~\eqref{NSys4}--\eqref{NSys6} can be formally treated as a consequence of the bulk Eqs.~\eqref{NSys1}--\eqref{NSys3} and an additional condition
\begin{equation}\label{Beta0}
\beta_{0n}=0,\quad (n\ge 1)\:.
\end{equation}
 We use the following ansatz for the doublon edge states  satisfying the requirement Eq.~\eqref{Beta0}:
\begin{multline}\label{EdgeStatesAnsatz}
\beta_{mn}=A\,\left[e^{ik_1 m}-e^{-ik_1 m}\right]\,e^{ik_2 n}\\\equiv
2i\,A\sin (k_1m)\,e^{ik_2n}\mspace{15mu} (m\leq n-1)\:.
\end{multline}
 The coefficients $\beta_{mm}$ are defined by Eqs.~\eqref{NSys1}, \eqref{NSys4}, and we also use the symmetry  property $\beta_{mn}=\beta_{nm}$. Equation \eqref{NSys3} yields the law of dispersion of two-photon excitations [cf. Eq.~\eqref{DispLaw1}]:
\begin{equation}\label{DispGeneral}
\eps=-2J\,\left(\cos\,k_1+\cos\,k_2\right)\:.
\end{equation}
Excluding $\beta_{mm}$ from Eqs.~\eqref{NSys1}, \eqref{NSys2}, we obtain

\begin{align}\label{Substitution}
[\left(\eps-2W\right)&(\eps-2U)-4J^2]\,\beta_{m,m+1)}\\\notag-
&2\,J^2\,[\beta_{(m-1,m}+\beta_{m+1,m+2)}]\\
+&J\,(\eps-2U)\,[\beta_{m,m+2}+\beta_{m-1,m+1}]=0\:.\notag
\end{align}
We plug the ansatz Eq.~\eqref{EdgeStatesAnsatz} into the Eq.~\eqref{Substitution} and derive two equations for both $e^{ik_1m+ik_2n}$ and $e^{-ik_1m+ik_2n}$ terms. After some algebra we find the following equations that determine the energy and the degree of localization of doublon edge states
\begin{gather}
\eps=-2J\left(\cos k_1+\cos k_2\right)\:,\label{EdgeEq1}\\
\eps-2U=-4iJ\sin k_2\:,\label{EdgeEq2}\\
\eps=2W(1-e^{2ik_2})\:.\label{EdgeEq3}
\end{gather}
The solutions of the system Eqs.~\eqref{EdgeEq1}--\eqref{EdgeEq3} should satisfy additional constraints: either $|e^{ik_2}|<|e^{ik_1}|<1$ or $|e^{ik_2}|<|e^{-ik_1}|<1$, otherwise the doublon state is not localized.

\begin{figure}[b]
\begin{center}
\includegraphics[width=1.0\linewidth]{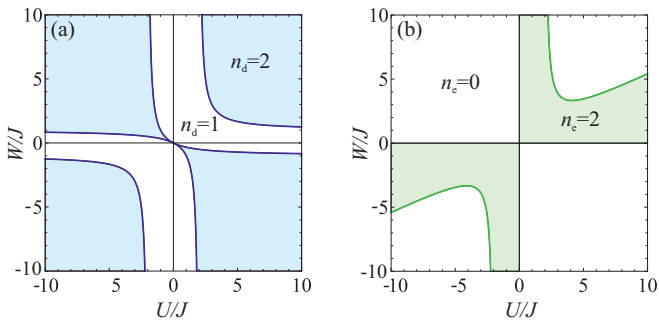}
\caption{Phase diagrams of (a) stable doublon branches and (b) doublon edge states. (a) Parameter domains corresponding to the two stable doublon branches are highlighted by blue. (b) Parameter domains corresponding to the two doublon edge states are shown by green.} 
\label{fig:FullDiagram}
\end{center}
\end{figure}

Given the analytical results Eqs.~\eqref{EdgeEq1}--\eqref{EdgeEq3}, it is straightforward to plot the phase diagram showing the range of existence of doublon edge states. To this end we notice that the condition for the edge state collapse reads $k_1=\pm k_2$ since in this case the long-range tail of the $\beta_{mn}$ distribution Eq.~\eqref{EdgeStatesAnsatz} acquires an infinite length. Further analysis of Eqs.~\eqref{EdgeEq1}--\eqref{EdgeEq3} shows that the edge state collapse occurs for the following relation between $U$ and $W$:
\begin{equation}\label{Wcoll}
W=W_c\equiv \frac{U}{2}\,\frac{U^2+4\,J^2}{U^2-4\,J^2}\:.
\end{equation}
In such critical scenario the localization parameter is equal to $e^{ik_2}=-2J/U$, and the energy of doublon edge state reads $\eps=(U^2+4\,J^2)/U$. In the limit $U\gg J$ the edge state collapses for $W_c\approx U/2$ in agreement with the approximate result of Sec.~\ref{sec:EffHam}, Eq.~\eqref{Cond}.

Figure~\ref{fig:FullDiagram}(b) illustrates the domain of existence of doublon edge states. In our analysis performed for a semi-infinite array a single edge state is observed. However, a finite array supports two edge states at both  edges. For different signs of $U$ and $W$ parameters, edge states do not exist regardless of the stability of doublon bands.

\begin{figure}[b]
\begin{center}
\includegraphics[width=1.0\linewidth]{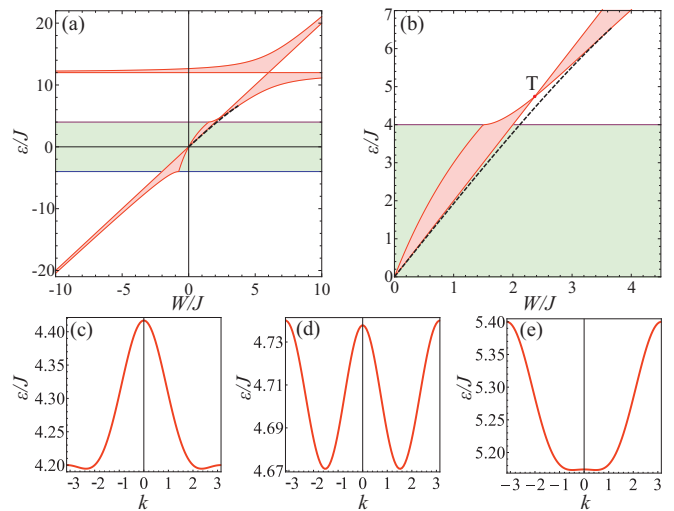}
\caption{(a,b) Energy bands of quasi-independent photons (green) and doublons (red) versus $W/J$ for the fixed value of $U/J=6$. Dashed line indicates the energy of the doublon edge state. Panel  (b)  shows the relative positions of the doublon W-band and the doublon edge state in more detail. Point T indicates the transition from negative to positive doublon mass. (c-e) Dispersion of doublon $W$-band for the fixed value $U/J=6$ and (c) $W/J=2.1$; (d) $W/J=2.37$; (e) $W/J=2.7$.} 
\label{fig:EWDiagram}
\end{center}
\end{figure}

It is also instructive to trace the evolution of  doublon bands for the fixed value of $U$ and varying magnitude of $W$ [Fig.~\ref{fig:EWDiagram}(a)]. In the region of negative $W \lesssim -2\,J$ both doublon branches are stable being separated from each other by the continuum of quasi-independent photon states. In this domain the system does not support edge states in agreement with the diagram Fig.~\ref{fig:FullDiagram}(b). In the region $-2\,J\lesssim W\lesssim 2\,J$ the doublon $W$-band is subject to collapse due to the interaction with the continuum of quasi-independent photon states which affects the band width. For $W\approx 2.37\,J$ the  $W$-band becomes flat. This can be explained by the mutual cancellation of the two terms in the tunneling amplitude $f=\tau+\rho$, discussed in the previous Section, that takes place at $W=U/3$ in the regime of strong interaction $W,U\gg J$. As the effective tunneling amplitude passes through zero as a function $W$,  the sign of the doublon effective mass  changes, as shown by the dispersion curves in Figs.~\ref{fig:EWDiagram}(c-e).  Doublon edge state exists only in a sufficiently narrow region of $W$ values and has quite small spectral separation from the doublon $W$-band [Fig.~\ref{fig:EWDiagram}(b)]. It should be highlighted that the doublon edge state is stable with respect to the interaction with the continuum of quasi-independent photon states since the decay of edge state into quasi-independent photons is prohibited by the momentum conservation.


\section{Conclusions}\label{sec:Concl}

We have analyzed two-photon states in a one-dimensional array of coupled nonlinear cavities described within extended Hubbard model. It has been shown that even though the edge resonator of the array is not detuned and the underlying one-particle model is topologically trivial, the system can still support edge states of bound photon pairs arising purely due to interactions. The composite nature of the  two-photon edge states is manifested in their bi-exponential wave function Eq.~\eqref{EdgeStatesAnsatz}.

Our studies reveal various transitions between the two-photon states including the decay of bulk doublons into quasi-independent photons (doublon collapse) or sign change of the doublon effective mass. Contrary to the bulk doublons, doublon edge states decay neither to the single-photon edge states nor to quasi-independent photons thus exhibiting certain robustness.

An additional feature of this model is the possibility to excite edge states at {\it both} edges of the chain simultaneously. We thus envision that the edge states formed by strongly correlated photons provide a rich potential for quantum information applications as well as quantum precision measurements with NOON states.

\begin{acknowledgments}
This work was supported by the Russian Science Foundation (Grant No.~16-19-10538).
A.N.P.  acknowledges the support by the Russian President Grant No.
MK-8500.2016.2. 
\end{acknowledgments}

\appendix
\section{Derivation of the doublon effective Hamiltonian}\label{sec:AppEffHam}
To get a qualitative understanding of the doublon edge state emergence in the system, we deduce the effective single-band doublon Hamiltonian. To this end we truncate the equations Eqs.~\eqref{Sys1}--\eqref{Sys3} leaving only the close-to-diagonal coefficients $\beta_{mn}$ with $|m-n|\leq 2$:
\begin{gather}
(\eps-2U)\,\beta_{mm}=-2J\,\beta_{m,m+1)}-2J\,\beta_{m-1,m}\mspace{15mu} (m\geq 2)\:,\label{Simpl1}\\
(\eps-2W)\,\beta_{m,m+1}=-J\,\left[\beta_{m+1,m+1}+\beta_{m,m+2}\right.\notag\\
+\left.\beta_{mm}+\beta_{m-1,m+1)}\right]\mspace{15mu} (m\geq 1)\:,\label{Simpl2}\\
\eps\,\beta_{m,m+2}=-J\,\left[\beta_{m+1,m+2}+\beta_{m,m+1}\right]\mspace{15mu} (m\geq 1)\:,\label{Simpl3}
\end{gather}

\begin{align}
(\eps-2U)\,&\beta_{11}=-2J\,\beta_{12}\:,\label{Simpl1b}\\
(\eps-2W)\,&\beta_{12}=-J\,\left(\beta_{11}+\beta_{22}+\beta_{13}\right)\:,\label{Simpl2b}\\
\eps\,&\beta_{13}=-J\,\left(\beta_{12}+\beta_{23}\right)\:.\label{Simpl3b}
\end{align}
First we analyze the doublon $U$-band. For such analysis it is sufficient to take into account only the coefficients $\beta_{mn}$ with $|m-n|\leq 1$ (two-diagonal approximation). We arrive to the relations:
\begin{align}
\eps\,&\alpha_1=(2U+j)\,\alpha_1+j\,\alpha_2\:,\label{UB1}\\
\eps\,&\alpha_n=(2U+2j)\,\alpha_n+j(\alpha_{n-1}+\alpha_{n+1})\mspace{15mu}(n\geq 2)\label{UB2}\:,
\end{align}
where $j=2J^2/(\eps-2W)\approx J^2/(U-W)$. Equations~\eqref{UB1}--\eqref{UB2} describe the one-dimensional chain with the tunneling constant $j$ and the detuning of the edge resonator $-j$. The edge states appear only when the detuning exceeds the tunneling constant \cite{Gorlach}, thus, the doublon edge states  are impossible for this band. Indeed, numerical calculations and exact analytical model do not reveal any doublon edge states in the vicinity of the $U$-band.

Second, we study the properties of the $W$-band within three-diagonal approximation. In this analysis we exclude the $\beta_{mm}$ and $\beta_{m(m+2)}$ coefficients leaving only the terms of the form $\beta_{m(m+1)}\equiv\eta_m$. The equations for the $\eta_m$ coefficients read:
\begin{align}
\eps\,\eta_1&=(2W+2\tau+\rho)\,\eta_1+(\tau+\rho)\,\eta_2\:,\label{WB1}\\
\eps\,\eta_n&=(2W+2\tau+2\rho)\,\eta_n\notag\\&+
(\tau+\rho)\,\left(\eta_{n-1}+\eta_{n+1}\right)\:,\mspace{15mu} (n\geq 2)\label{WB2}
\end{align}
where $\tau=2J^2/(\eps-2U)\approx J^2/(W-U)$ and $\rho=J^2/\eps\approx J^2/2W$. Equations \eqref{WB1}--\eqref{WB2} again describe a simple chain with the edge resonator detuning equal to $-\rho$ and demonstrate a perfect agreement with the perturbative treatment in Sec.~\ref{sec:EffHam}.

\bibliography{TopologicalLib2}

\end{document}